\documentstyle[12pt,epsf]{article}
\advance\voffset by -1.5cm
\advance\hoffset by -2.1cm
\def\be{\begin{equation}}
\def\ee{\end{equation}}

\def\Zop{\bbbz}
\def\Rop{\bbbr}
\def\Nop{\bbbn}
\def\pmb#1{\setbox0=\hbox{#1}
 \kern-.025em\copy0\kern-\wd0
 \kern.05em\copy0\kern-\wd0
 \kern-.025em\raise.0433em\box0 }

\def\Ct{C_{\!t}}
\def\I{{\cal I}}

\def\Tr{{\rm Tr}}

\def\RR{R$-$R }
\def\NSNS{NS$-$NS }
\def\dh{$\hat{\mbox{D}}$}

\def\ba              {\begin{eqnarray}}
\def\ea              {\end{eqnarray}}
\def\spc{\;\;\;\;}
\def\dh{$\hat{\mbox{D}}$}

\def\3{\ss}
\def\sq{\hbox{\rlap{$\sqcap$}$\sqcup$}}
\def\qed{\ifmmode\sq\else{\unskip\nobreak\hfil
\penalty50\hskip1em\null\nobreak\hfil\sq
\parfillskip=0pt\finalhyphendemerits=0\endgraf}\fi}
\def\half {\frac{1}{2}}

\def\bbbz {{\sf Z\!\!Z}}

\def\bbbr {{\rm I\!R}}
\def\bbbn {{\rm I\!N}}

\def\xxx#1           {{hep-th/#1}}

\def\npb#1(#2)#3     {Nucl. Phys. {\bf B#1} (#2) #3 }
\def\rep#1(#2)#3     {Phys. Rept.{\bf #1} (#2) #3 }
\def\plb#1(#2)#3     {Phys. Lett. {\bf #1B} (#2) #3 }
\def\prl#1(#2)#3     {Phys. Rev. Lett.{\bf #1} (#2) #3 }
\def\prd#1(#2)#3     {Phys. Rev. {\bf D#1} (#2) #3 }
\def\ap#1(#2)#3      {Ann. Phys. {\bf #1} (#2) #3 }
\def\rmp#1(#2)#3     {Rev. Mod. Phys. {\bf #1} (#2) #3 }
\def\cmp#1(#2)#3     {Comm. Math. Phys. {\bf #1} (#2) #3 }
\def\mpla#1(#2)#3    {Mod. Phys. Lett. {\bf A#1} (#2) #3 }
\def\ijmp#1(#2)#3    {Int. J. Mod. Phys. {\bf A#1} (#2) #3 }
\def\cqg#1(#2)#3     {Class. Quant. Grav. {\bf #1} (#2) #3 }
\def\am#1(#2)#3      {Adv. Math. {\bf #1} (#2) #3 }
\def\im#1(#2)#3      {Invent. Math. {\bf #1} (#2) #3 }
\def\jhep#1(#2)#3    {J. High Energy Phys. {\bf #1} (#2) #3 }

\newcommand{\ket}[1]{|#1\rangle}

\def\Tr{{\rm Tr}}

\def\ss{\bf S}

\def\Nu{{\cal N}}
\def\ie{{\it i.e.}}

\begin{document}

\thispagestyle{empty}
\def\thefootnote{\fnsymbol{footnote}}
\begin{flushright}
  hep-th/9910109\\
  DAMTP-1999-142
 \end{flushright}
\vskip 0.5cm

\begin{center}\LARGE
{\bf Dirichlet Branes on Orbifolds}
\end{center}
\vskip 1.0cm
\begin{center}
{\large  Matthias R Gaberdiel\footnote{E-mail  address: 
{\tt M.R.Gaberdiel@damtp.cam.ac.uk}}} and 
{\large  Bogdan Stefa\'nski, jr.\footnote{E-mail  address: 
{\tt B.Stefanski@damtp.cam.ac.uk}}} 

\vskip 0.5 cm
{\it Department of Applied Mathematics and Theoretical Physics \\
Centre for Mathematical Sciences \\
Wilberforce Road \\
Cambridge CB3 0WA, U.K.}
\end{center}

\vskip 1.0cm

\begin{center}
October 1999
\end{center}

\vskip 1.0cm

\begin{abstract}
The D-brane spectrum of a class of orbifolds of toroidally
compactified Type IIA and Type IIB string theory is analysed
systematically. The corresponding K-theory groups are determined 
and complete agreement is found. The charge densities of the various
branes are also calculated.
\end{abstract}

\vskip 1.0cm 
\begin{center}
\end{center}

\vfill
\setcounter{footnote}{0}
\def\thefootnote{\arabic{footnote}}
\newpage

\renewcommand{\theequation}{\thesection.\arabic
{equation}}

\section{Introduction}
\setcounter{equation}{0}

It is now generally appreciated that Dirichlet branes (D-branes)
\cite{Dai,mbg1,Pol1} play a central r\^ole in the non-perturbative
description of string theory. D-branes are solitonic solutions of the
underlying supergravity theory but they have also a description in
terms of open strings; this allows for an essentially perturbative
treatment of D-branes.

The D-branes that were first analysed were BPS states that break half
the (spacetime) supersymmetry. It has now been realised however that,
because of their description in terms of open strings, D-branes can be
constructed and analysed in much more general situations. In fact,
D-branes are essentially described by a boundary conformal field
theory \cite{PolCai,CLNY,Cardy,Lew}, the consistency conditions of
which are not related to spacetime supersymmetry \cite{BG1,BG2,KT}. In
an independent development, D-branes that break supersymmetry have
been constructed in terms of bound states of branes and anti-branes
by Sen \cite{Sen1,Sen2,Sen3,Sen4,Sen5,Sen6}. This construction has
been interpreted in terms of K-theory by Witten \cite{WittenK}, and
this has opened the way for a more mathematical treatment of D-branes 
\cite{Horava,Gukov,BGH}.  

The D-brane spectrum of a number of theories is understood in
detail. These include the standard ten-dimensional Type IIA, IIB and I
theory (see \cite{polv2} for a review and \cite{Sen4,Sen5,Frau0} for
more recent developments), as well as their non-supersymmetric
cousins, Type 0A, 0B and 0 \cite{BG1,KT,BG4}. It is understood how the
D-brane spectrum is modified upon compactification on tori, some
supersymmetric orbifolds and near ALE singularities
\cite{dm,Frau1,Frau2,Sen6,BG3,DDG,DG,GabSen}. There has also been
progress in understanding the D-brane spectrum of Gepner models
\cite{RS1,GS} and WZW theories \cite{PSS1,PSS2,AS,ARS,FFFS}.   

In this paper we analyse systematically the D-brane spectrum of
certain orbifolds of toroidal compactifications of IIA/IIB
superstring theory\footnote{This class of compactifications contains 
supersymmetric orbifolds such as the ones analysed in
\cite{Sen6,BG3,GabSen}, but also non-supersymmetric theories.}. We 
describe in detail the boundary states that define the different
D-branes, and show that the resulting spectrum is in agreement with 
the K-theory predictions which we determine independently. 
We also find the charge densities of the branes, and exhibit (for a
number of examples) the difference between the K-theory group
describing the D-brane charges and cohomology.  
Our analysis works equally well for supersymmetric theories as well as
for theories without supersymmetry, thus emphasising once again that
D-branes do not intrinsically depend on supersymmetry.
\smallskip

The paper is organised as follows. In section~2, we describe the
perturbative and non-perturbative (D-brane) spectrum of the theories
in question. Section~3 gives a brief account of the K-theory
calculation in the uncompactified case. This is modified in section~4 
to take into account the effect of the toroidal compactification. In
section~5 we determine the charge densities, and section~6 contains
some concluding remarks. We have included three appendices where some
of the more technical details can be found.

\section{The Dirichlet brane spectrum}
\setcounter{equation}{0} 

In the first subsection we describe briefly the spectrum of the
various orbifold theories we shall be considering. In subsection
\ref{sect22} we then identify the consistent boundary states which
form the building blocks for the Dirichlet branes.

\subsection{Orbifold theories}\label{sect21}

Let us consider the orbifold of Type IIA or Type IIB that is generated
by the non-trivial generator 
\be
\label{orbifold} 
g_1 = \I_n \qquad \hbox{or} \qquad g_2 = \I_n (-1)^{F_L} \,.
\ee
Here $\I_n$ describes the reflection of $n$ coordinates, and
$(-1)^{F_L}$ acts as $\pm 1$ on left-moving spacetime bosons and
fermions, respectively. The action of $g_i$ describes a symmetry of
Type II theories if $n$ is even, and we shall therefore only consider
this case. In the first instance we shall assume that the theory is
ten-dimensional, but later on (in Sections~\ref{sect4}
and~\ref{sect5}) we shall also discuss how the analysis is modified if
the $n$ directions on which $\I_n$ acts are compactified on an
$n$-torus. In this case, T-duality relates  
\be\label{Tdu1}
\hbox{IIA}\; \hbox{on}\; T^n / \I_n  \longleftrightarrow 
\hbox{IIB}\; \hbox{on}\; T^n / \I_n (-1)^{F_L}\,,
\ee
and similarly 
\be\label{Tdu2}
\hbox{IIA}\; \hbox{on}\; T^n / \I_n (-1)^{F_L}  \longleftrightarrow 
\hbox{IIB}\; \hbox{on}\; T^n / \I_n \,.
\ee
The orbifold group is $\Zop_2$ if $n=4$ mod $4$, and in this case the
theory is supersymmetric. For $n=2$ mod $4$, $g_1^2=g_2^2=(-1)^{F^s}$,
where $F^s$ is the spacetime fermion number, and the orbifold is
actually $\Zop_4$. In this case supersymmetry is broken since the
orbifold theory does not contain any spacetime fermions. The orbifold
of Type IIA/IIB by $(-1)^{F^s}$ is Type 0A/0B, and thus the $\Zop_4$
orbifold can equivalently be described as the $\Zop_2$ orbifold of
Type 0A/0B by $g_1$ or $g_2$; this is the point of view we shall
adopt. 

For definiteness we shall phrase our results for the supersymmetric
theories in the following, but we shall explain, where appropriate,
the modifications that arise for $n=2$ mod $4$. In particular, we
shall analyse carefully the construction of the boundary state
components for all theories (see appendix~B). Given the results of
\cite{BG1,KT,BG4} it is then easy to determine the actual D-brane
spectrum from this data. In fact the only modification that occurs is
that the D-brane spectrum of the Type 0A/0B theories is doubled
compared to that of Type IIA/IIB. Since the K-theory analysis is also
doubled (as there are two different D9-branes in Type 0A and Type 0B),
this is then also in agreement with the K-theory analysis.    

Let us start by describing the bosonic (closed string) spectrum of
these theories that is relevant for the description of the boundary
states. In the untwisted sector there are the states in the \NSNS and
the \RR sectors that are invariant under the orbifold projection 
$\half(1+g)$ (where $g=g_1$ or $g=g_2$), and the GSO-projection  
\be
\label{GSO}
\begin{array}{ll}
\hbox{NS-NS} & {1 \over 4} \left(1+(-1)^F\right) 
\left(1 + (-1)^{\tilde{F}}\right) \vspace*{0.1cm} \\
\hbox{\RR} & {1 \over 4} \left(1+(-1)^F\right) 
\left(1 \mp (-1)^{\tilde{F}}\right) \,,
\end{array}
\ee
where, in the second line, the $-$ sign corresponds to Type IIA and
the $+$ sign to Type IIB. In the twisted sector, the moding for the
excitations corresponding to the $n$ directions along which $\I_n$ acts
is half-integral for bosons and the R-sector world-sheet fermions,
and integral for the NS-sector world-sheet fermions. Furthermore, the
ground state energy vanishes in the twisted R-sector, and is 
\be
a_{\scriptsize{\hbox{NS,T}}} = {(n-4) \over 8} 
\ee
in the twisted NS-sector. If the orbifold projection does not involve
$(-1)^{F_L}$, \ie\ if $g=g_1$, then the GSO-projection in the twisted
sector is the same as (\ref{GSO}). On the other hand, for $g=g_2$, the
left-moving GSO-projection is opposite to (\ref{GSO}) in all twisted
sectors. Furthermore, the states in the twisted sector also have to be
invariant under the orbifold symmetry.

The lowest lying states in the twisted \RR sector are always massless,
and transform as a tensor product of two spinor representations of
$SO(8-n)$ (where $SO(8-n)$ acts on the unreflected $8-n$ coordinates
in light-cone gauge); this tensor product can be decomposed into
antisymmetric tensor representations of $SO(8-n)$. The orbifold
projection is very simple in this case since the twisted R-sector does
not have any fermionic zero modes along the $n$ directions of
$\I_n$. All massless (GSO-invariant) states of the twisted \RR sector
are therefore physical.

In the twisted NS-NS sector, the lowest lying states are massless for
$n=4$, massive for $n>4$, and tachyonic for $n=2$. 
They transform as a tensor product of two spinor representations of
$SO(n)$ (where $SO(n)$ acts on the $n$ coordinates that are reflected
by $\I_n$); 
from the point of view of the $8-n$ unreflected directions these
states are scalars. On the ground states, the orbifold projection is 
proportional to the GSO-projection, and all GSO-invariant ground
states are again physical.

We should mention at this stage that the definition of $\I_n$ in the
untwisted R-R sector is in general ambiguous: on the ground states we
can either define $\I_n$ to be
\be
\I^{(1)}_n = \prod_{\mu=9-n}^{8} (\sqrt{2}\psi^\mu_0) 
       \prod_{\mu=9-n}^{8} (\sqrt{2} \tilde\psi^\mu_0)
\ee
or
\be
\I^{(2)}_n = \prod_{\mu=9-n}^{8}  (2 \psi^\mu_0 \tilde\psi^\mu_0) \,.
\ee
These two expressions differ by the order of the fermionic zero modes;
for $n=4$ mod (4), $\I^{(1)}_n= \I^{(2)}_n$ but for $n=2$ mod (4), 
$\I^{(1)}_n= -\I^{(2)}_n$. In this paper we shall use the convention
that $\I_n$ refers to $\I^{(1)}_n$; for $n=2$ mod (4), we then have
$\I_n (-1)^{F_L} = \I^{(2)}_n$.

For $n=2$ mod $4$, there is furthermore, at least {\it a priori},
an ambiguity in how to define the GSO-projection in each of the
twisted sectors; this is due to the fact that in order to guarantee
that $(-1)^F$ and $(-1)^{\tilde{F}}$ are of order two, one has to
introduce non-trivial phases which are only determined up to signs. As
we shall explain in appendix~B, these ambiguities are uniquely fixed if
we require that the theory has at least one `fractional' D-brane. This
convention leads to a D-brane spectrum that is consistent with the
results obtained independently using K-theory.

\subsection{The boundary state analysis}\label{sect22}

Let us first introduce a convenient notation to describe the
allignement of the D-branes relative to the $n$ preferred
directions along which $\I_n$ acts. For definiteness, let us assume
that $\I_n$ reflects the coordinates $x^{9-n},\ldots,x^8$; we then say
that a Dirichlet $p$-brane is of type $(r,s)$ where $p=r+s$ if it has
$r+1$ Neumann directions along $x^0,\ldots, x^{8-n}, x^9$, and $s$
Neumann directions along $x^{9-n},\ldots,x^8$. We shall always work in
light-cone gauge (with light-cone directions $x^0$ and $x^9$), and
therefore the actual D-brane states we shall analyse will have
Dirichlet boundary conditions along the two light cone directions, and
thus be D-instantons. As usual, we shall assume that we can perform an
appropriate Wick-rotation to transform these states back to ordinary
D-branes \cite{GrGut}.  

The analysis we shall now describe is very similar to that performed
in \cite{BG1,Sen2,BG2} (see also \cite{PolCai,CLNY}), and we shall
therefore be rather sketchy. (Some details of the construction can
however be found in appendix~A.) A D-brane is described by a 
linear combination of physical boundary states that satisfies a
certain compatibility condition. For each set of boundary conditions,
there exists at most one non-trivial GSO and orbifold invariant
boundary state in each sector (untwisted \NSNS and \RR, and twisted
\NSNS and \RR), which is unique up to normalisation. The
compatibility condition requires that the spectrum of states
that is induced by the presence of a collection of D-branes defines 
open strings that have consistent interactions with the original
closed string theory. These open strings can be determined, using
world-sheet duality, from the corresponding closed string tree
diagrams that describes the exchange  of a closed string state between
two D-branes.  

One aspect of this consistency condition is the requirement that the
string that begins and ends on the same D-brane must be a suitably
projected open string. This implies that the actual D-brane state
consists of non-trivial boundary states from different sectors. In
fact, there exist three different possibilities: the boundary state
describes either a {\em fractional} D-brane, \ie\ it is a linear
combination  of non-trivial boundary states from all four sectors,
\be
\label{fractional}
\ket{D(r,s)} = {1\over2} \left(\ket{D(r,s)}_{\mbox{\scriptsize\NSNS}} 
+ \varepsilon_1 \ket{D(r,s)}_{\mbox{\scriptsize\RR}}
+ \varepsilon_1 \varepsilon_2 \ket{D(r,s)}_{\mbox{\scriptsize\NSNS,T}} 
+ \varepsilon_2 \ket{D(r,s)}_{\mbox{\scriptsize\RR,T}} \right)\,,
\ee
a {\it bulk} D-brane, \ie\ it is a linear combination
involving only the untwisted \NSNS and \RR sectors
\be
\label{bulk}
\ket{D(r,s)}_b = \ket{D(r,s)}_{\mbox{\scriptsize\NSNS}} +
\varepsilon \ket{D(r,s)}_{\mbox{\scriptsize\RR}}\,, 
\ee
or a {\it truncated} D-brane, \ie\ it is a linear combination
involving only the untwisted \NSNS and the twisted \RR sector,
\be
\label{nonBPS}
\ket{\hat{D}(r,s)} = {\Nu\over\sqrt{2}}
\left(\ket{D(r,s)}_{\mbox{\scriptsize\NSNS}} 
+ \varepsilon \ket{D(r,s)}_{\mbox{\scriptsize\RR,T}} \right)\,.
\ee
Here $\varepsilon$ and $\varepsilon_{1,2}$ are $\pm 1$ and describe
the signs with respect to the different charges, the relative
normalisations are determined by the consistency condition, and $\Nu$
is a normalisation constant that will be determined further below. 
The first two D-brane states (\ref{fractional}) and (\ref{bulk})
are conventional D-branes that are BPS provided that the
orbifold preserves supersymmetry, whereas the state in (\ref{nonBPS})
describes a non-BPS brane.

Not all of these D-brane states are independent: two
fractional D-branes with $\varepsilon_1=\varepsilon_1'$ and
$\varepsilon_2=-\varepsilon_2'$ can combine to form a bulk D-brane
state (\ref{bulk}); as it turns out, whenever a bulk D-brane exists,
then so does the corresponding fractional brane, and we may thus
restrict ourselves, without loss of generality, to considering only
fractional and truncated D-brane states. 

The condition that a string with both endpoints on the same truncated
brane defines a consistent open string requires that $\Nu^2\in\Nop$; 
the minimal value is therefore $\Nu=1$. On the other hand, the
compatibility condition also requires that a string which begins on
any one of these branes and ends on any other one, must describe an
open string. If for a given $(r,s)$, a fractional D-brane state
exists, then the open string from the fractional $(r,s)$ to the
truncated $(r,s)$ D-brane leads to the partition function 
\be
\sqrt{2} \Nu (\mbox{NS} - \mbox{R}) \half \left(1 + g (-1)^F \right) \,.
\ee
This only defines an open string partition function if $\Nu$ is an
integer multiple of $\sqrt{2}$, and the minimal value is then
$\Nu=\sqrt{2}$. In this case, the mass and the charge of the 
truncated D-brane is precisely twice that of the fractional D-brane;
thus the truncated D-brane can decay into two fractional D-branes with
$\varepsilon_1=-\varepsilon_1'$ and $\varepsilon_2=\varepsilon_2'$,
and does not describe an independent stable state. 

In summary we therefore find that for a given $(r,s)$ at most
one of the above three D-brane states is fundamental, and that
the other two (if they exist) can be obtained as bound states of the
fundamental D-brane. This fundamental D-brane is either fractional
or truncated. The fundamental D-brane is only stable provided that
the open string that begins and ends on it does not have a tachyon.
This is always the case for a fractional D-brane; in the case of
a truncated D-brane the stability of the brane depends on the actual
value of the compactification radii. In fact, if the theory is
uncompactified, a truncated D-brane is unstable if and only if $s>0$,
and in the compactified case it is stable provided that the radii of
the tangential circles are sufficiently small (and the radii of the
circles transverse to the brane are sufficiently large)
\cite{Sen6,BG3,GabSen}. 

The D-brane spectrum can now be determined by analysing which of the
different boundary components are GSO- and orbifold invariant. The
detailed analysis is described in appendix~B, and the final result is \\

\begin{tabular}{cll} 
(i) & {\bf IIA by $\I_n$}: & Fractional $(r,s)$ D-branes exist for 
$r$ and $s$ both even. \\
& & Truncated $(r,s)$ D-branes exist for $r$ even and  $s$ odd. \\
(ii) & {\bf IIB by $\I_n$}: & Fractional $(r,s)$ D-branes exist for 
$r$ odd and $s$ even. \\
& & Truncated $(r,s)$ D-branes exist for both $r$ and $s$ odd. 
\end{tabular}
\\

\begin{tabular}{cll} 
(iii) &  {\bf IIA by $\I_n(-1)^{F_L}$}: & Fractional $(r,s)$ D-branes
exist for $r$ and $s$ both odd. \\
& & Truncated $(r,s)$ D-branes exist for $r$ odd and $s$ even. \\
(iv) & {\bf IIB by $\I_n (-1)^{F_L}$}: & Fractional $(r,s)$ D-branes
exist for $r$ even and $s$ odd. \\
& & Truncated $(r,s)$ D-branes exist for both $r$ and $s$ even.
\end{tabular}

\noindent In the uncompactified theory, a fractional D-brane has two
charges ($\varepsilon_1$ and $\varepsilon_2$), and a truncated D-brane
has only a single charge; the corresponding K-theory groups should
therefore either be $\Zop\oplus\Zop$ or $\Zop$, depending on $(r,s)$
as above. Strictly speaking, one should only trust this analysis for
$s=0$, since the truncated D-brane is otherwise unstable. However, as
we shall see in the next section, the K-theory result agrees with the
above even for $s\ne 0$.

The situation is somewhat clearer in the case where all $n$ directions 
are compactified; this will be discussed in some detail in
section~4.

\section{K-theory analysis in the uncompactified theory}
\setcounter{equation}{0}

In this section we demonstrate that the above results can be
reproduced in terms of K-theory. Here we shall only consider the
uncompactified theory; the compactified case will be discussed in
section~4. We shall discuss Type IIB in some detail in the first
subsection, and Type IIA is analysed in section~3.2. 

\subsection{K-theory of Type IIB}

The analysis of the D-brane spectrum in Type IIB orbifolds is
relatively straightforward. If the $\Zop_2$ orbifold leaves the
D9-brane invariant ({\it i.e.} if it is of type $g=g_1$), the D-brane
spectrum is described in terms of equivariant K-theory 
$K_{\Zop_2}$~\cite{Segal} of the space transverse to the worldvolume of the
Dirichlet brane; on the other hand, if the orbifold maps
the D9-brane to its anti-brane ({\it i.e.} if it is of
type $g=g_2$), the relevant K-theory group is 
$K_{\pm}$~\cite{WittenK}. For a given $(r,s)$ brane, the transverse space has
dimension $9-(r+s)$, of which $n-s$ directions are inverted under the
action of $\I_n$. In order to distinguish between the directions on
which $\I_n$ does or does not act, we denote the transverse space (as
in \cite{Gukov}) by $\Rop^{n-s,9-n-r}$. The K-theory groups we want to
determine are then 
\be
K_{\Zop_2}(\Rop^{n-s,9-n-r}) \qquad\hbox{and}\qquad
K_{\pm}(\Rop^{n-s,9-n-r}) \,.
\ee
The D-brane configurations of interest are equivalent to the vacuum
at transverse infinity; in terms of K-theory this means that the pairs
of  bundles $(E,F)$ that define the elements of K-theory have the
property  that $E$ is isomorphic to $F$ near infinity
\cite{WittenK}. The corresponding K-theory is usually called K-theory
{\em with compact support}. 
By a theorem of Hopkins, $K_{\pm}(\Rop^{l,m})$ is given as
\be
\label{kpm}
K_\pm(\Rop^{l,m})=K^1_{\Zop_2}(\Rop^{l+1,m})=K_{\Zop_2}(\Rop^{l+1,m+1})\,,
\ee
where the last equality follows from the compact support condition. 
The calculation of $K_{\pm}(\Rop^{l,m})$ therefore reduces to that of
the equivariant K-theory groups. This has been calculated
before by Gukov~\cite{Gukov}, and the result is 
\be
\label{Kresult}
K_{\Zop_2}(\Rop^{l,m})=
\left\{\begin{array}{ll}
0 &  \hbox{if $m$ is odd}, \\
\Zop &  \hbox{if $m$ is even and $l$ is odd}, \\
R[\Zop_2]=\Zop\oplus\Zop & \hbox{if $m$ and $l$ are even,}
\end{array}\right.
\ee
where $R[G]$ is the representation ring of the group $G$. This is
precisely in agreement with the results of the boundary analysis that
we described in the previous section. 

We shall now give a different derivation of (\ref{Kresult}) that is
due to Segal \cite{Segal1}. Firstly, because of Bott periodicity, the
answer depends only on the parity of $l$ and $m$. If $l$ is even we
have   
\be
\label{equivariant}
K_{\Zop_2}(\Rop^{2\hat{l},m})=K_{\Zop_2}(\Rop^{0,m})
=\left\{\begin{array}{ll}
0 & \hbox{if $m$ is odd,} \\
R[\Zop_2]=\Zop\oplus\Zop & \hbox{if $m$ is even,} \\
\end{array}\right.
\ee
where we have used the fact that $\Zop_2$ acts freely on
$\Rop^{0,m}$. In order to calculate $K_{\Zop_2}(\Rop^{1,m})$ we
observe that  
\be\label{eq1}
K^*_{\Zop_2}(X\times\Rop^{1,0})
=K^*_{\Zop_2}(X\times D^1,X\times S^0)\,,
\ee
where $X$ is an arbitrary manifold on which $\Zop_2$ acts continuously, and 
$S^0\subset D^1\subset\Rop^{1,0}$ are the one-dimensional `circle' 
({\it i.e.} the two points $\pm 1$) and the one-dimensional `disk'
({\it i.e.} the interval $[-1,1]$), respectively. The group $\Zop_2$
acts on $S^0$ and $D^1$ by reflection. Because of (\ref{eq1}), the
long exact sequence can be written as 
\ba
\cdots\rightarrow K^{-1}_{\Zop_2}(X\times S^0) &
\rightarrow & K^0_{\Zop_2}(X\times \Rop^{1,0}) \;
\rightarrow \; K^0_{\Zop_2}(X\times D^1)\nonumber \\
& \rightarrow & K^0_{\Zop_2}(X\times S^0) \;
\rightarrow \; K^1_{\Zop_2}(X\times \Rop^{1,0}) \;
\rightarrow \; K^1_{\Zop_2}(X\times D^1) \;
\rightarrow\cdots\label{les}
\ea
By homotopy equivalence, we have
$K^*_{\Zop_2}(X\times D^1) \cong  K^*_{\Zop_2}(X)$, and 
$K^*_{\Zop_2}(X\times S^0)\cong K^*(X)$ since the $\Zop_2$ action is
free. Thus for $X=\Rop^{l,m}$ with $l$ and $m$ even, the exact
sequence becomes  
\be
0\rightarrow K^0_{\Zop_2}(\Rop^{l,m}\times \Rop^{1,0})
 \rightarrow R[\Zop_2]\rightarrow \Zop
 \rightarrow K^1_{\Zop_2}(\Rop^{l,m}\times \Rop^{1,0})\rightarrow 0\,,
\ee
where we have used (\ref{equivariant}), $K^0(\Rop^{l,m})=\Zop$ and 
$K^1(\Rop^{l,m})=0$. The map from $R(\Zop_2)=\Zop\oplus\Zop$ to $\Zop$
is given by $(m,n)\mapsto m+n$ (which is onto), and it follows that 
\begin{eqnarray}
K^0_{\Zop_2}(\Rop^{l,m}\times\Rop^{1,0}) & = & \Zop\,, \\
K^1_{\Zop_2}(\Rop^{l,m}\times\Rop^{1,0}) & = & 0\,.
\end{eqnarray}
Thus for $l=m=0$ we obtain the desired result.

\subsection{Orbifolds of Type IIA}

The D-brane spectrum of Type IIA theory is given in terms of
the K-theory groups $K^{-1}(X)$ \cite{WittenK,Horava}. The elements of
$K^{-1}(X)$ can be thought of as pairs $(E,\alpha)$, where $E$
is a bundle on $X$, and $\alpha$ a bundle-automorphism \cite{Karoubi}.
In terms of string theory, the D-branes of Type IIA theory can be
obtained from (unstable) D9-branes by tachyon condensation. This can
then be interpreted directly in terms of the pairs $(E,\alpha)$,
where $E$ is the bundle on the D9-branes, and the automorphism $\alpha$ 
is related to the tachyon field $T$ by
\be
\alpha=-e^{\pi i T}\,.
\ee
This construction also applies to the orbifold theories in question,
except that now the tachyon field has to be invariant under the
orbifold action. In terms of K-theory, the relevant group is then the
equivariant group $K_{\Zop_2}^1(X)$ that consists of pairs
$(E,\alpha)$, where $\alpha$ commutes with the $\Zop_2$ action
\cite{Segal1}. More precisely, the K-theory groups are 
\be
K^{-1}_{\Zop_2}(\Rop^{n-s,9-n-r}) \qquad\hbox{and}\qquad
K^{-1}_{\pm}(\Rop^{n-s,9-n-r}) \,,
\ee
where again the first case corresponds to $g_1$, and the second to 
$g_2$ orbifolds. As before $\Rop^{n-s,9-n-r}$ denotes the transverse
space to a $(r,s)$-brane. Because of the compact support condition,
$K_\pm^{-1}$ can be evaluated using equation~(\ref{kpm}). Explicitly
this gives  
\be
K^{-1}_{\Zop_2}(\Rop^{n-s,9-n-r})=K_{\Zop_2}(\Rop^{n-s,10-n-r}) 
\qquad\hbox{and}\qquad
K^{-1}_{\pm}(\Rop^{n-s,9-n-r})=K_{\pm}(\Rop^{n-s,10-n-r}) \,,
\ee
and thus the evaluation of K-groups relevant to the Type IIA orbifolds
under consideration follows from equation~(\ref{Kresult}), and is in
precise agreement with the results of the boundary state analysis
described in the previous section.

\section{The compactified case}\label{sect4}
\setcounter{equation}{0}

In this section we analyse the D-brane spectrum for the case when all
$n$ coordinates along which $\I_n$ acts are compactified on a
torus. We shall first describe how the K-theory analysis is modified;
we shall then explain how this matches precisely the results that 
can be obtained using the boundary state formalism.

\subsection{Relative K-theory}

The D-brane spectrum of toroidally compactified Type II string 
theories can be described in terms of relative K-theory
\cite{BGH}, and the relevant K-theory groups are therefore
\be 
K_{\Zop_2}^*(S^{9-n-r}\times T^n,T^n) \qquad\hbox{and}\qquad
K_\pm^*(S^{9-n-r}\times T^n,T^n)\,.
\ee 
Here the involution acts on $T^n$ with $2^n$ fixed points and has
trivial action on $S^{9-n-r}$, where $S^{9-n-r}$ is the one-point
compactification of the space $\Rop^{9-n-r}$ transverse to the brane
in the uncompactified directions. The D-branes of interest are trivial
at infinity, and the K-theory groups are therefore the
relative K-theory groups that describe pairs of bundles $(E,F)$ where
$E$ and $F$ are isomorphic at $\{\infty\}\times T^n$. The K-theory
groups only classify branes with fixed $r$; in terms of our previous
discussion, these K-theory groups therefore correspond to products
over different values of $s$.

Let us first compute the equivariant relative K-theory groups. Since
$T^n$ is a retract of $S^{9-n-r}\times T^n$ we have
\be
K^*_{\Zop_2}(S^{9-n-r}\times T^n,T^n)\oplus
K^*_{\Zop_2}(T^n)=K^*_{\Zop_2}(S^{9-n-r}\times T^n)\,.
\ee
Here, we may take $K^*_{\Zop_2}(T^n)$ and
$K^*_{\Zop_2}(S^{9-n-r}\times T^n)$ to denote the unreduced equivariant
K-theory groups. The former have been computed \cite{Segal1}
\ba
K_{\Zop_2}(T^n)&=&3 \cdot 2^{n-1}\;\Zop\,, \label{t1}\\ 
K^1_{\Zop_2}(T^n)&=&0\,,\label{t2}
\ea
where we use the notation that $n \Zop\equiv \Zop^{\oplus n}$. 
Furthermore we have 
\ba
K^*_{\Zop_2}(X\times S^{2k})&=&K^*_{\Zop_2}(X)\oplus K^*_{\Zop_2}(X)\,,
\label{evensplit} \\ 
K^*_{\Zop_2}(X\times S^{2k+1})&=&K^*_{\Zop_2}(X)\oplus K^{*-1}_{\Zop_2}(X)\,.
\label{oddsplit}
\ea
Taking these identities together and using the fact that $n$ is even
we then find that
\ba
K_{\Zop_2}(S^{9-n-r}\times T^n,T^n)&=&\left\{\begin{array}{ll}
3 \cdot 2^{n-1} \Zop&\mbox{$r$ odd,} \\ 
0&\mbox{$r$ even,}\end{array}\right. \label{47}
\\
K^1_{\Zop_2}(S^{9-n-r}\times
T^n,T^n)&=&\left\{\begin{array}{ll} 3 \cdot 2^{n-1}\Zop
&\mbox{$r$ even,} \\ 0&\mbox{$r$ odd.}\end{array}\right. \label{48}
\ea
Next we consider $K_{\pm}$. As before we have
\be
K^*_{\pm}(S^{9-n-r}\times T^n,T^n)\oplus
\tilde{K}^*_{\pm}(T^n)=\tilde{K}^*_{\pm}(S^{9-n-r}\times T^n)\,,
\ee
which we have now written in terms of the reduced
$\tilde{K}$-groups. Because of the theorem of Hopkins we have  
\ba
\tilde{K}^m_{\pm}(T^n)&=&K^{m-1}_{\Zop_2}(T^{n}\times\Rop^{1,0})\,,\\
\tilde{K}^m_{\pm}(S^{9-n-r}\times T^n) & = & 
K^{m-1}_{\Zop_2}(S^{9-n-r}\times T^{n}\times\Rop^{1,0})\,.
\ea
Using (\ref{evensplit}) and (\ref{oddsplit}) this implies that 
\be
K^m_{\pm}(S^{9-n-r}\times T^n,T^n)=\left\{\begin{array}{ll} 
K_{\Zop_2}^{m+1}(T^n\times\Rop^{1,0})
&\mbox{$r$ odd,} \\ K_{\Zop_2}^m(T^n\times\Rop^{1,0})
&\mbox{$r$ even}\,.\end{array}\right. 
\ee
To compute $K_{\Zop_2}^*(T^n\times\Rop^{1,0})$ we consider next the
long exact sequence (\ref{les}) with $X=T^n$. Using (\ref{t2}) this
becomes
\be\label{eq3}
0\rightarrow K^{-1}(T^n)\rightarrow 
K_{\Zop_2}(T^n\times \Rop^{1,0}) \rightarrow 
\tilde{K}_{\Zop_2}(T^n) \rightarrow 
\tilde{K}(T^n)\rightarrow 
K_{\Zop_2}^1(T^n\times \Rop^{1,0})\rightarrow 0\,,
\ee 
where we have observed that for compact manifolds K-theory with
compact support is the same as reduced K-theory. It is easy to see
that (\ref{eq3}) remains true if we replace the two reduced
$\tilde{K}$-groups by their corresponding unreduced groups. We can
then use (\ref{t1}) together with $K^*(T^n)=2^{n-1}\Zop$ to rewrite
this as 
\be\label{eq4}
0\rightarrow 2^{n-1}\; \Zop \stackrel{\alpha}{\rightarrow} 
K_{\Zop_2}(T^n\times \Rop^{1,0})\stackrel{\beta}{\rightarrow} 
3 \cdot 2^{n-1}\; \Zop \stackrel{\gamma}{\rightarrow} 
2^{n-1}\; \Zop \rightarrow K_{\Zop_2}^1(T^n\times \Rop^{1,0})
\rightarrow 0\,.
\ee 
Since $\beta$ is injective and the forgetting map $\gamma$ is onto we
then find that 
\begin{eqnarray}
K_{\Zop_2}(T^n\times\Rop^{1,0}) & = & 3\cdot 2^{n-1}\; \Zop\,,\\
K^1_{\Zop_2}(T^n\times \Rop^{1,0}) & = & 0\,.
\end{eqnarray}
Finally, we thus have
\ba
K_\pm^{-1}(S^{9-n-r}\times T^n,T^n)&=&\left\{\begin{array}{ll}
3 \cdot 2^{n-1} \Zop&\mbox{$r$ odd,} \\ 
0&\mbox{$r$ even,}\end{array}\right. \\
K_\pm(S^{9-n-r}\times
T^n,T^n)&=&\left\{\begin{array}{ll} 3 \cdot 2^{n-1}\Zop
&\mbox{$r$ even,} \\ 0&\mbox{$r$ odd.}\end{array}\right.
\ea
In agreement with T-duality (\ref{Tdu1}), (\ref{Tdu2}), these K-theory
groups are the same as (\ref{47}) and (\ref{48}), respectively. 

\subsection{Compact boundary states}

Boundary states for D-branes that extend along compact (and inverted)
directions have been analysed before \cite{Sen2,GabSen}. In essence
the boundary states are described by the same formulae that we have
given in section~\ref{sect22} and appendix~\ref{appb1}; there are
however a few (minor) differences. Firstly, the branes can wind along
the internal directions of the torus, and therefore carry appropriate
winding numbers. Secondly, the orbifold has $2^n$ fixed points in the
compactified case, and there are therefore $2^n$ different twisted
sectors. For $s>0$ the D-branes extend along the internal directions,
and the corresponding boundary states have a contribution from $2^s$
of these twisted sectors. (In fact, the $2^s$ twisted sectors
correspond to the $2^s$ endpoints of the $s$-dimensional world-volume
of the brane along the internal directions.) The structure of the
boundary states, and in particular the normalisation of the different
boundary components, is described in detail in appendix~A.

In all known examples \cite{SenRev} (see also section~5), the only
charges that are conserved in the various decay processes among
D-branes are the R-R charges in the untwisted and twisted sectors. As
we have just seen, there are $2^n$ different twisted R-R sectors; if
for a given $r$, the corresponding twisted R-R boundary state is
allowed, there are then $2^n$ different twisted R-R sector charges. As
regards the charges in the untwisted R-R sector, these arise from the 
10-dimensional R-R forms upon compactification on the different cycles
of the torus. If the orbifold is of type $g_1$, the relevant cycles
are even-dimensional, and we therefore get
\be
N_{1} = \sum_{\stackrel{l=0}{l\; \mbox{\scriptsize{even}}}}^{n}
{n\choose l} = 2^{n-1} \,, 
\ee
whereas for $g_2$ we have
\be
N_2 = \sum_{\stackrel{l=1}{l\; \mbox{\scriptsize{odd}}}}^{n-1} 
{n\choose l} = 2^{n-1} \,.
\ee
It follows from the analysis in section 2 and appendix~B that the
condition on $r$ for the twisted R-R sector boundary component to be
consistent is the same as that for the untwisted R-R sector
component. Thus, if $r$ satisfies this condition, there are altogether 
\be
N= 2^n + 2^{n-1} = 3 \cdot 2^{n-1}
\ee
R-R charges that form a lattice of dimension $3 \cdot 2^{n-1}$ (and
otherwise there are none). Since these are the only charges that are
conserved in transitions between different D-branes, the actual
D-brane charges form then a sublattice of this lattice. 

In general this sublattice is not the whole lattice (see in particular 
appendix~C for a concrete example), but it is of maximal rank; this
follows from the fact that it contains yet another sublattice, namely
the lattice of D-brane charges that is generated by  the bulk and the
truncated D-branes. (This is clearly of maximal rank since the bulk
D-branes are only charged under the untwisted R-R sector, whereas the
truncated D-branes are only charged under the different twisted R-R
sectors.) It therefore follows that the lattice of D-brane charges is
$3 \cdot 2^{n-1}\Zop$ if $r$ satisfies the appropriate condition, and
zero otherwise. This is consistent with the result that follows from
K-theory.

\section{Charge densities}\label{sect5}
\setcounter{equation}{0}

In this section we determine the charge densities of the different 
D-branes that we have described in this paper. We shall from now on
always consider the compactified case; the descent relations apply
equally to the Type 0A/0B case, but the overall normalisation is
slightly different in that case.

\subsection{Descent relations}

As we have mentioned before, various D-branes can decay into one
another. The basic phenomenon from which all others can be obtained is
that of a truncated D-brane \dh$(r,s)$ decaying into two fractional
D$(r,s')$-branes with $s'=s+1$ or into two fractional D$(r,s')$-branes
with $s'=s-1$. Since the R-R sector charges are conserved by these
processes, this implies certain relations between the charges of the
different 
D-branes. In order to obtain these we introduce the following
notation. Let us label the $2^n$ fixed points by $1,\ldots, 2^n$. A
$(r,s)$ D-brane is charged under $2^s$ of these $2^n$ fixed points,
and we include their labels as suffices, \ie\ the $(r,s)$ brane that
is charged under the fixed points $n_1,\ldots,n_{2^s}$ is denoted by 
\be\label{labels} 
(r,s)_{n_1,\ldots,n_{2^s}}\,.
\ee
If the charge with respect to one of the twisted R-R sectors is
opposite, we place a bar over the corresponding label; similarly, if
a fractional D-brane has opposite untwisted R-R sector charge, we
place a bar over $s$ in (\ref{labels}). 

The basic decay processes can now be described as follows: a truncated 
\dh$(r,s)$ brane can decay into two fractional D-branes with $s'=s+1$ as 
\be
\hbox{\dh}(r,s)_{n_1,\ldots,n_{2^s}} \longrightarrow 
\hbox{D}(r,s+1)_{n_1,\ldots,n_{2^s},m_1,\ldots,m_{2^s}} + 
\hbox{D}(r,\overline{s+1})_{n_1,\ldots,n_{2^s},
    \overline{m_1},\ldots,\overline{m_{2^s}}} \,,
\ee
or it can decay into two fractional D-branes with $s'=s-1$ as
\be
\hbox{\dh}(r,s)_{n_1,\ldots,n_{2^s}} \longrightarrow 
\hbox{D}(r,s-1)_{n_1,\ldots,n_{2^{s-1}}} + 
\hbox{D}(r,\overline{s-1})_{n_{2^{s-1}+1},\ldots,n_{2^s}}\,.
\ee
It follows from this observation that the twisted R-R sector charge of
a fractional D-brane D$(r,s+1)$ at each fixed point is {\em half} that
of a truncated D-brane \dh$(r,s)$, and that the twisted R-R sector
charge of a fractional D-brane D($r,s-1)$ at each fixed point is the
{\em same} as that of a truncated D-brane \dh$(r,s)$. We can apply
this argument repeatedly to express the charge of any D-brane in terms 
of that of the brane with $s=0$. There are two cases to consider: if
the orbifold is of type $g_1$, the $(r,0)$ brane is fractional, and we 
have 
\be
\begin{array}{lrcl}
\mbox{D}(r,2k) \quad & 
\mu'_{(r,2k)} & = & 2^{-k}\mu^\prime_{(r,0)}\,, \\
\mbox{\dh}(r,2k+1) \qquad &
\mu'_{(r,2k+1)} & = & 2^{-k}\mu^\prime_{(r,0)}\,,
\end{array}
\ee
where $\mu'_{(r,s)}$ denotes the twisted R-R charge density of the
a D$(r,s)$-brane. Similarly, if the orbifold of type $g_2$, the
$(r,0)$ brane is truncated, and we find instead
\be
\begin{array}{lrcl}
\mbox{D}(r,2k+1)\quad &
\mu'_{(r,2k+1)} & = & 2^{-k-1}\mu^\prime_{(r,0)^\prime}\,, \\
\mbox{\dh}(r,2k)\quad &
\mu'_{(r,2k)} & = & 2^{-k}\mu^\prime_{(r,0)^\prime}\,.
\end{array}
\ee
These relations can also be obtained from the normalisation constants
of the various branes that are determined in appendix~A; see in
particular Eq.~(\ref{fracnormtw}), (\ref{truncnormtw}).

\subsection{The overall normalisation}\label{candt} 

The above considerations only determine the twisted charges up to an
overall factor. In order to find this normalisation constant we shall
now compare the (open) string theory calculation with a field theory
calculation \cite{Pol1}. We shall only consider the compactified case;
because of the arguments of section~5.1 it is then sufficient to do
the calculation for a brane with $s=0$. 

If the orbifold theory is of type $g_1$, the $(r,0)$ brane is
fractional, and its open string has the partition function 
\be\label{fracopen}
\int\frac{dt}{2t}\Tr_{NS-R}\left[\frac{1}{4}(1+(-1)^F)
(1+\I_n)e^{-2tH_o} \right]\,.
\ee
The twisted \RR sector contribution comes from NS$(-1)^F \I_n$ and
can be evaluated as
\be
\label{a}
-\frac{1}{4}\frac{V_{r+1}}{(2\pi)^{r+1}}\int\frac{dt}{t}
(4\pi\alpha^\prime t)^{-(r+1)/2}
e^{-t\frac{Y^2}{2\pi\alpha^\prime}}
\frac{f_4^{8-n}(\tilde{q})f_3^{n}(\tilde{q})}
{f_1^{8-n}(\tilde{q})2^{-n/2}f_2^{n}(\tilde{q})}\,,
\ee
where $Y$ is the separation of the two branes, $V_{r+1}$ denotes the
(infinite) world-volume area of the $(r,0)$-brane, and $q$,
$\tilde{q}$ and the functions $f_i$ are defined as in appendix~A.  In
the field theory limit $(t\rightarrow 0)$ this gives  
\ba
-\frac{1}{4}\frac{V_{r+1}} {(2\pi)^{r+1}}& &\!\!\!\!\!\!\!\!\!\!\!\!
\int\frac{dt}{t} (4\pi\alpha^\prime t)^{-(r+1)/2}
e^{-t\frac{Y^2}{2\pi\alpha^\prime}}
t^{(8-n)/2}(16+O(e^{-\pi/t}))\nonumber \\
&\simeq&-V_{r+1} \pi(4\pi^2\alpha^\prime)^{3-r}G_{9-r}(Y)\,.
\ea
where $G_{d}$ is the scalar Green's function in $d$ dimensions 
({\it c.f.} Eq.~(13.3.2) of \cite{polv2}). This is to be compared with
the field theory calculation, where the relevant terms in the
effective action are
\be
-\frac{1}{4\kappa_{10-n}^2}\int d^{10-n}x\;\sqrt{g}
(H_{\!t}^{(r+2)})^2+\mu^\prime_{(r,0)}\int \Ct^{(r+1)}\,.
\ee
The field theory amplitude is
\be
-2 \left(\mu'_{(r,0)} \kappa_{10-n} \right)^2 G_{9-r}(Y)\,,
\ee
and comparison of the two results then gives
\be
(\kappa_{10-n}\mu^\prime_{(r,0)})^2
=\half\pi(4\pi^2\alpha^\prime)^{3-r}\,.
\ee
It follows from the arguments of section~5.1 that the general formula
for the twisted R-R charge density $\mu^\prime_{(r,s)}$ of a $(r,s)$
brane on an $\I_n$ orbifold is then  
\be
(\kappa_{10-n}\mu^\prime_{(r,s)})^2=
2^{-(2k+1)}\pi(4\pi^2\alpha^\prime)^{3-r}\,,
\ee
where $s=2k$ or $s=2k+1$ for fractional and truncated branes,
respectively. 

In the case of the $g_2$ orbifold the $(r,0)$ brane is truncated. The 
partition function of the relevant open string is then
\be
\int\frac{dt}{2t}\Tr_{NS-R}\left[\frac{1}{2}(1+(-1)^F \I_n)
e^{-2tH_o}\right]\,.
\ee
The twisted \RR charge contribution comes again from the
NS$(-1)^F\I_n$ sector and is exactly twice that in
equation~(\ref{a}). On the other hand, the field theory action is as
before, and we therefore find  
\be
(\kappa_{10-n}\mu_{(r,0)^\prime}^\prime)^2
=\pi(4\pi^2\alpha^\prime)^{3-r}\,.
\ee
Denoting by $\mu^\prime_{(r,s)^\prime}$ the twisted \RR charge
density of an $(r,s)$-brane on a $(-1)^{F_L}\I_n$ orbifold we thus have
\be
(\kappa_{10-n}\mu^\prime_{(r,s)^\prime})^2=
2^{-2k}\pi(4\pi^2\alpha^\prime)^{3-r}\,,
\ee
where now $s=2k$ or $s=2k-1$ for a truncated or fractional brane,
respectively.

By similar methods one can also determine the charge densities with
respect to the untwisted \RR sector, and the tensions of the different
branes.

\section{Conclusions}
\setcounter{equation}{0}

In this paper we have analysed systematically the D-branes of certain
orbifolds of (toroidal compactifications) of Type IIA/IIB string
theory. We have determined the corresponding K-theory groups, and we
have found complete agreement with the results obtained from a
boundary state analysis. We have also calculated the relevant 
charge densities. 

It would be interesting to determine the world-volume actions of the
branes we have considered in this paper; work in this direction is in
progress \cite{Ste}.

\appendix

\section{Construction and normalisation of boundary states}\label{appb}
\setcounter{equation}{0}

In this appendix we determine the normalisation constants of the
boundary states for the orbifold theories under consideration. We
shall use the conventions of \cite{Sen2,GabSen}. Let us first consider
the uncompactified case.  

\subsection{The uncompactified case}\label{appb1}

In each (bosonic) sector of the theory we can construct the boundary
state 
\be\label{boundary1}
\ket{B(r,s),k,\eta}=\exp\left(\sum_{l>0}^\infty 
\left[\frac{1}{l}\alpha_{-l}^\mu S_{\mu\nu}\tilde{\alpha}_{-l}^\nu\right]
+i\eta\sum_{m>0}^\infty \left[\psi_{-m}^\mu
S_{\mu\nu}\tilde{\psi}_{-m}^\nu\right]\right)\ket{B(r,s),k,\eta}^{(0)}\,,
\label{bdr}
\ee
where, depending on the sector, $l$ and $m$ are integer or
half-integer, and $k$ denotes the momentum of the ground state.  We
shall always work in light-cone gauge with light-cone directions $x^0$
and $x^9$; thus $\mu$ and $\nu$ take the values $1,\ldots, 8$. We
shall also drop the dependence on $\alpha^\prime$ from now on.

The parameter $\eta=\pm$ describes the two different spin structures 
\cite{PolCai,CLNY}, and the matrix $S$ encodes the boundary conditions
of the Dp-brane which we shall always take to be diagonal
\be
S=\mbox{diag}(-1,\dots,-1,1,\dots,1) \,,
\ee
where $p+1$ entries are equal to $-1$, $7-p$ entries are equal to
$+1$, and $p=r+s$. If there are fermionic zero modes, the ground state
in (\ref{boundary1}) satisfies an additional condition. (This will be 
relevant in appendix~B; see for example (\ref{bound}) and
(\ref{bound1}).) 

In order to obtain a localised D-brane, we have to take the Fourier
transform of the above boundary state, where we integrate over the
directions transverse to the brane,
\be\label{local}
\ket{B(r,s),y,\eta}=\int\left(\prod_{\mu \;\mbox{\scriptsize{transverse}}} 
dk^\mu e^{ik^\mu y_\mu}\right) dk^0e^{ik^0y_0} \;
dk^9 e^{ik^9 y_9} \; \ket{B(r,s),k,\eta}\,,
\ee
$y$ denotes the location of the boundary state, and in the twisted
sectors the momentum integral only involves transverse directions that
are not inverted by the orbifold action. In the following we shall
always consider (without loss of generality) the case $y=0$ in which
case the boundary state is denoted by $\ket{B(r,s),\eta}$.  

The invariance of the boundary state under the GSO-projection always
requires that the physical boundary state is a linear combination of
the two states corresponding to $\eta=\pm$. Using the conventions of
appendix~B, these linear combinations are of the form
\ba
\left|B(r,s)\right>_{\mbox{\scriptsize\NSNS}}&=&
\frac{1}{2}\Bigl(\left|B(r,s),+\right>_{\mbox{\scriptsize\NSNS}}-
\left|B(r,s),-\right>_{\mbox{\scriptsize\NSNS}}\Big)\,,\\
\left|B(r,s)\right>_{\mbox{\scriptsize\RR}}&=&
\frac{4i}{2}\Bigl(\left|B(r,s),+\right>_{\mbox{\scriptsize\RR}}+
\left|B(r,s),-\right>_{\mbox{\scriptsize\RR}}\Bigr)\,,\\ 
\left|B(r,s)\right>_{\mbox{\scriptsize\NSNS,T}}&=&
\frac{2^{n/4}}{2}
\Bigl(\left|B(r,s),+\right>_{\mbox{\scriptsize\NSNS,T}}
+\left|B(r,s),-\right>_{\mbox{\scriptsize\NSNS,T}}\Bigr)\,,\\
\left|B(r,s)\right>_{\mbox{\scriptsize\RR,T}}&=&
\frac{2^{2-n/4}i}{2}
\Bigl(\left|B(r,s),+\right>_{\mbox{\scriptsize\RR,T}}
+\left|B(r,s),-\right>_{\mbox{\scriptsize\RR,T}}\Bigr)\,,\label{rrtw}
\ea
where, depending on the theory in question, these states are actually 
GSO-invariant provided that $r$ and $s$ satisfy suitable conditions. 
The normalisation constants have been introduced for later
convenience. 

In order to solve the open-closed consistency condition the actual
D-brane state is a linear combination of physical boundary states from 
different sectors. There are two cases to consider, fractional and
truncated D-branes. In the former case, the D-brane state can be
written as 
\begin{eqnarray}
\ket{D(r,s)} & = & {\cal N}_{f,U}
\left(\left|B(r,s)\right>_{\mbox{\scriptsize\NSNS}} 
+ \epsilon_1 \left|B(r,s)\right>_{\mbox{\scriptsize\RR}} \right) 
\nonumber \\
& & \quad 
+ \epsilon_2 \; {\cal N}_{f,T}
\left(\left|B(r,s)\right>_{\mbox{\scriptsize\NSNS,T}} 
+ \epsilon_1 \left|B(r,s)\right>_{\mbox{\scriptsize\RR,T}} \right)\,,
\end{eqnarray}
where $\epsilon_i=\pm$ determines the sign of the charge with respect
to the untwisted and twisted \RR sector charge. The closed string
cylinder diagram is then of the form\footnote{The minus sign in the term
proportional to $\delta_{8,n}$ does not, at first sight, agree with the
conventions of equation~(\ref{rrtw}). However for $n=8$ the twisted
\RR sector does not have any zero modes, and the ground state is
therefore unique; since the orbifold preserves supersymmetry
\cite{dasmuk}, the sign of the GSO projection in this case is then
determined by supersymmetry.}
 \ba
\label{closedcal}
{\cal A} & = & \int dl \left<B(r,s)\right|e^{-lH_c}\left|B(r,s)\right>
\nonumber \\ 
& = & \half {\cal N}^2_{f,U} \int dl \; l^{(p-9)/2}
\left(\frac{f_3^8(q)-f_4^8(q)-f_2^8(q)}{f_1^8(q)}\right) \nonumber \\
& & + \half {\cal N}^2_{f,T} \int dl \; l^{(r+n-9)/2}
\left(\frac{f_3^{8-n}(q)f_2^n(q)-
f_2^{8-n}(q)f_3^n(q)-\delta_{8,n}f_4^8(q)}{f_1^{8-n}(q)f_4^n(q)}\right) 
\,,
\ea
where the functions $f_i$ are defined as in \cite{PolCai}, 
$q=e^{-2\pi l}$, and the closed string Hamiltonian is given by  
\be
H_c=\pi k^2+2\pi\sum_{\mu=1}^8\left[\sum_{l>0}^\infty
(\alpha^\mu_{-l}\alpha^\mu_l+\tilde{\alpha}^\mu_{-l}\tilde{\alpha}^\mu_l)+
\sum_{m>0}^\infty
m(\psi^\mu_{-m}\psi^\mu_m+\tilde{\psi}^\mu_{-m}\tilde{\psi}^\mu_m)\right]
+2\pi C_c\,.
\ee
Here the constant $C_c$ is $-1$ in the \NSNS sector, zero in the
untwisted and twisted \RR sector, and $(n-4)/4$ in the twisted \NSNS
sector. The corresponding open string amplitude is obtained by the 
modular transformation $t=1/2l$, ${\tilde q}=e^{-\pi t}$,
\ba
{\cal A} & = & 
2^{(7-p)/2} {\cal N}^2_{f,U} \int {dt \over 2t}
t^{-(p+1)/2} 
\left(\frac{f_3^8(\tilde{q})-f_2^8(\tilde{q})
-f_4^8(\tilde{q})}{f_1^8(\tilde{q})}\right) \nonumber \\
& & + 2^{(7-n-r)/2} {\cal N}^2_{f,T} \int {dt \over 2t}
t^{-(r+1)/2} \left(\frac{f_3^{8-n}({\tilde q})f_4^n({\tilde q})-
f_4^{8-n}({\tilde q})f_3^n({\tilde q})-\delta_{8,n}
f_2^8({\tilde q})}
{f_1^{8-n}({\tilde q}) f_2^n({\tilde q})}\right)\,.\label{theclosedone}
\ea
This is to be compared with the open string one-loop diagram,
\ba
\int\frac{dt}{2t}&
&\!\!\!\!\!\!\!\!\!\!\!\!\!\!\!
\Tr_{NS-R}\left(\frac{1+(-1)^F}{2}\frac{1+g}{2}
e^{-2tH_o}\right)\nonumber \\
&=&
\frac{V_{p+1}}{(2\pi)^{p+1}}2^{-(p+5)/2}\int
\frac{dt}{2t}t^{-(p+1)/2}
\left(\frac{f_3^8({\tilde q})-f_4^8({\tilde q})-
f_2^8({\tilde q})}{f_1^8({\tilde q})}\right)
\nonumber \\
& &+\frac{V_{r+1}}{(2\pi)^{r+1}}2^{(n-r-5)/2}\int
\frac{dt}{2t}t^{-(r+1)/2}
\left(\frac{f_3^{8-n}({\tilde q})f_4^n({\tilde q})
-f_4^{8-n}({\tilde q})f_3^n({\tilde q})-\delta_{8,n}f_2^8({\tilde q})}
{f_1^{8-n}({\tilde q})f_2^n({\tilde q})}\right)\,,\nonumber\\
\label{theopenone}
\ea
where $g$ denotes the orbifold operator, $V_{p+1}$ is the
(infinite) $p+1$ dimensional volume of the brane, whilst
$V_{r+1}$ is the volume of the projection onto the directions
unaffected by $\I_n$. The open string Hamiltonian is given by
\be
H_o=\pi p^2+\pi\sum_{\mu=1}^8\left[\sum_{l>0}^\infty \alpha_{-l}^\mu 
\alpha_l^\mu+\sum_{m>0}^\infty m\psi^\mu_{-m}\psi^\mu_m\right] +\pi
C_o\,, 
\ee
where, in the R sector, $l$ and $m$ run over the positive integers for
NN and DD directions, and over positive half integers for ND
directions. In the NS sector, the moding of the fermions (and
therefore the values for $m$) are opposite to those in the R sector.
$C_o$ is zero in the R sector and is $\frac{4-t}{8}$ in the NS sector,
where $t$ is the number of ND directions. Comparison of
equations~(\ref{theopenone}) and~(\ref{theclosedone}) then gives
\ba
{\cal N}^2_{f,U}&=&\frac{V_{p+1}}{(2\pi)^{p+1}}
{1\over 64}\,, \\
{\cal N}^2_{f,T}&=&\frac{V_{r+1}}{(2\pi)^{r+1}}
{2^{n} \over 64}\,.\label{con1}
\ea
The analysis for the case of the truncated D-branes is similar. 
In this case the D-brane boundary state is given by
\be
\ket{\hat{D}(r,s)} = {\cal N}_{t,U}
\left|B(r,s)\right>_{\mbox{\scriptsize\NSNS}} 
+ \epsilon {\cal N}_{f,T}
\left|B(r,s)\right>_{\mbox{\scriptsize\RR,T}}\,,
\ee
where $\epsilon=\pm$ determines the sign of the twisted \RR sector
charge. The closed string tree diagram now only produces some of the
terms of (\ref{closedcal}), and the corresponding open string
amplitude is 
\ba
\int\frac{dt}{2t}&
&\!\!\!\!\!\!\!\!\!\!\!\!\!\!\!
\Tr_{NS-R}\left(\frac{1+ g (-1)^F}{2}
e^{-2tH_o}\right)\nonumber \\
&=&
\frac{V_{p+1}}{(2\pi)^{p+1}}2^{-(p+3)/2}\int
\frac{dt}{2t}t^{-(p+1)/2}
\left(\frac{f_3^8({\tilde q})-f_2^8({\tilde q})}
{f_1^8({\tilde q})}\right) 
\nonumber \\
& & - \frac{V_{r+1}}{(2\pi)^{r+1}}2^{(n-r-3)/2}\int
\frac{dt}{2t}t^{-(r+1)/2}
\left(\frac{f_4^{8-n}({\tilde q})f_3^n({\tilde q})
-\delta_{8,n}f_2^8({\tilde q})}
{f_1^{8-n}({\tilde q})f_2^n({\tilde q})}\right)\,.\\
\ea
Comparison with the corresponding closed string calculation then gives
\ba
{\cal N}^2_{t,U}&=&\frac{V_{p+1}}{(2\pi)^{p+1}}{1 \over 32}\,, \\
{\cal N}^2_{t,T}&=&\frac{V_{r+1}}{(2\pi)^{r+1}}{2^{n}\over
32}\,. \label{con2} 
\ea

\subsection{The compactified case}

The construction in the compactified case is essentially the same as
in the above uncompactified case; however there are the following
differences. 

\noindent 1. In the localised boundary state (\ref{local}) the
integral over compact transverse directions is replaced by a sum
\be
\int dk^\nu e^{ik^\nu y_\nu} \longrightarrow
\sum_{m^\nu\in\Zop} e^{i m^\nu y_\nu / R_\nu} \,,
\ee
where $R_\nu$ is the radius of the compact $x^\nu$ direction.

\noindent 2. In the two untwisted sectors, the ground state is
in addition characterised by a winding number $w_\nu$ for each compact 
direction that is tangential to the world-volume of the brane. The
localised bound state (\ref{local}) then also contains a sum over these
winding states 
\be
\sum_{w_\mu} e^{i\theta^\mu w_\mu} \,,
\ee
where $\theta^\mu$ is a Wilson line; as required by orbifold
invariance, $\theta^\mu \in\{ 0,\pi\}$. 

\noindent 3. For general $s$, the contribution in the two twisted
sectors consists of a sum of terms that are associated to $2^s$ of the 
$2^n$ different twisted sectors that define the endpoints of the
world-volume of the brane in the internal space. For convenience we
may assume that one of the $2^s$ fixed points is always the origin.   

\noindent 4. The open and closed string Hamiltonians, $H_o$ and $H_c$,
each acquire an extra term $1/4\pi(\sum_\mu w_\mu^2)$.

Let us now construct in more detail the boundary state for a
fractional D$(r,s)$ brane. This is of the form
\begin{eqnarray}\label{fractboun}
\ket{D(r,s)} & = & {\cal N}_{f,U}
\left(\left|B(r,s)\right>_{\mbox{\scriptsize\NSNS}} 
+ \epsilon_1 \left|B(r,s)\right>_{\mbox{\scriptsize\RR}} \right) 
\nonumber \\
& & \quad 
+ \epsilon_2 \; {\cal N}_{f,T} \sum_{\alpha=1}^{2^s}
e^{i\theta_\alpha}
\left(\left|B(r,s)\right>_{\mbox{\scriptsize\NSNS,T$_\alpha$}} 
+ \epsilon_1 
\left|B(r,s)\right>_{\mbox{\scriptsize\RR,T$_\alpha$}} \right)\,,
\end{eqnarray}
where $\alpha$ labels the different fixed points between which the
brane stretches (where we choose the convention that $T_1$ is the
twisted sector at the origin), and $\theta_\alpha$ is the Wilson
line that is associated to the difference of the fixed point $\alpha$
and the origin. 
(Thus if $\alpha$ has coordinates $(n_i \pi R_i)$, $i=9-n,\ldots,8$,
$\theta_\alpha=n_i \theta^i$.) 
The closed string tree diagram is now
\ba
{\cal A}_c &=&  \int dl \left<B(r,s)\right|e^{-lH_c}\left|B(r,s)\right>
\nonumber \\ 
& = & \half{\cal N}^2_{f,U} \int dl \;
l^{(r+n-9)/2}
\left(\frac{f_3^8(q)-f_2^8(q)-f_4^8(q)}{f_1^8(q)}\right)
\nonumber \\
& & \qquad \qquad\times\;\;\
\prod_{i=1}^{s}\sum_{w_{j_i}\in\Zop} e^{-l\pi R_{j_i}^2 w_{j_i}^2}
\prod_{i=1}^{n-s}\sum_{n_{k_i}\in\Zop} e^{-l\pi (n_{k_i}/R_{k_i})^2}
\nonumber \\
& & \quad + {2^s \over 2} {\cal N}^2_{f,T} \int dl \; l^{(r+n-9)/2}
\left(\frac{f_3^{8-n}(q)f_2^n(q)-
f_2^{8-n}(q)f_3^n(q)-\delta_{8,n}f_4^8(q)}{f_1^{8-n}(q)f_4^n(q)}\right) 
\,,
\ea
where $R_{j_i}$, $i=1,\ldots, s$ are the radii of the circles that are
tangential to the world-volume of the brane, and $R_{k_i}$, 
$i=1,\ldots, n-s$ are the radii of the directions transverse to the
brane. Upon the substitution $t=1/2l$, using the Poisson resummation
formula (see for example \cite{Sen2,GabSen}), this amplitude becomes
\ba
{\cal A}_c &=& {\cal N}^2_{f,U} 
{\prod_{i=1}^{n-s} R_{k_i} \over \prod_{i=1}^{s}  R_{j_i}} 
2^{(7-r)/2} \int {dt \over 2t} t^{-(r+1)/2} 
\left(\frac{f_3^8(\tilde{q})-f_2^8(\tilde{q})
-f_4^8(\tilde{q})}{f_1^8(\tilde{q})}\right) \nonumber \\
& & \qquad \qquad\times\;\;\
\prod_{i=1}^{s}\sum_{n_{j_i}\in\Zop} e^{-2t\pi n_{j_i}^2 / R_{j_i}^2}
\prod_{i=1}^{n-s}\sum_{w_{k_i}\in\Zop} e^{-2t\pi w_{k_i}^2 R_{k_i}^2}
\nonumber \\
& & \quad + 2^{(7-n-r)/2} 2^s {\cal N}^2_{f,T} \int {dt \over 2t}
t^{-(r+1)/2} \left(\frac{f_3^{8-n}({\tilde q})f_4^n({\tilde q})-
f_4^{8-n}({\tilde q})f_3^n({\tilde q})-\delta_{8,n}
f_2^8({\tilde q})}
{f_1^{8-n}({\tilde q}) f_2^n({\tilde q})}\right)\,. \nonumber
\ea
This is to be compared with the open string amplitude
\ba
\int\frac{dt}{2t}
\Tr_{NS-R}\left(\frac{1+(-1)^F}{2}\frac{1+g}{2} e^{-2tH_o}\right)
& = & \frac{V_{r+1}}{4 (2\pi)^{r+1}} 2^{-(r+1)/2}
\int\frac{dt}{2t}\! t^{-(r+1)/2}
\frac{f_3^8({\tilde{q}})-f_4^8({\tilde q})-
f_2^8({\tilde q})}{f_1^8({\tilde q})}
\nonumber \\
& &\qquad\times\;
\prod_{i=1}^{s}\sum_{n_{j_i}\in\Zop} e^{-2t\pi n_{j_i}^2 / R_{j_i}^2}
\prod_{i=1}^{n-s}\sum_{w_{k_i}\in\Zop} e^{-2t\pi w_{k_i}^2 R_{k_i}^2}
\nonumber \\
& &  + \frac{V_{r+1}}{(2\pi)^{r+1}}2^{(n-r-5)/2}\int
\frac{dt}{2t}t^{-(r+1)/2} \nonumber \\
& & \qquad \times
\frac{f_3^{8-n}({\tilde q})f_4^n({\tilde q})
-f_4^{8-n}({\tilde q})f_3^n({\tilde q})-\delta_{8,n}f_2^8({\tilde q})}
{f_1^{8-n}({\tilde q})f_2^n({\tilde q})}\,.\nonumber\\
\ea
By comparison this then fixes the normalisation constants as
\ba
{\cal N}^2_{f,U}&=&\frac{V_{r+1}}{(2\pi)^{r+1}} {1\over 64}
{\prod_{i=1}^{s}  R_{j_i} \over \prod_{i=1}^{n-s} R_{k_i}} \,,
\label{fracnormun}\\
{\cal N}^2_{f,T}&=&\frac{V_{r+1}}{(2\pi)^{r+1}} {2^{n-s} \over 64}\,.
\label{fracnormtw}
\ea

The analysis for the truncated D-branes is almost identical; the
boundary state is the truncation of (\ref{fractboun}) to the 
untwisted \NSNS and the twisted \RR sector; this then only depends on  
one parameter $\epsilon=\epsilon_1 \epsilon_2$ as well as the Wilson
lines $\theta_\alpha$. The open string amplitude contains also only 
the corresponding terms. Furthermore, since the projection operator is 
now $\half(1+g(-1)^F)$ each of the terms that appears is twice as
large as in the fractional case. This implies that the relevant
normalisation constants are given as 
\ba
{\cal N}^2_{t,U}&=&\frac{V_{r+1}}{(2\pi)^{r+1}} {1\over 32} 
{\prod_{i=1}^{s}  R_{j_i} \over \prod_{i=1}^{n-s} R_{k_i}} \,,
\label{truncnormun}\\
{\cal N}^2_{t,T}&=&\frac{V_{r+1}}{(2\pi)^{r+1}} {2^{n-s} \over 32}\,.
\label{truncnormtw}
\ea
We should mention that the normalisation constants of the
twisted sector boundary states (\ref{fracnormtw}) and
(\ref{truncnormtw}) in the compactified theory differ from the
corresponding normalisation constants in the uncompactified theory
(\ref{con1}) and (\ref{con2}) by a factor of $2^s$. This seems
contradictory since the twisted charge that is carried by a D-brane
should not depend on whether the directions transverse to the orbifold
plane are compact or not. Presumably this means that we can only trust
the boundary state calculation if all directions tangential to the
brane are compact; this is anyway necessary for the normalisation
constants in the untwisted sector to make sense (since they are
proportional to the volume of the brane).\footnote{We thank Ashoke Sen
for a discussion on this point.}

\section{Consistency conditions of boundary states}\label{appa}
\setcounter{equation}{0}

In this appendix we analyse which boundary states are GSO- and
orbifold invariant. Let us first consider the condition that
comes from GSO-invariance. It is well known that the untwisted
\NSNS component is invariant under the GSO-projection for all
$(r,s)$. In the untwisted \RR sector, the GSO-projection acts as in
the theory before orbifolding, and therefore the boundary state is
only GSO-invariant if $r+s$ is even in the case of a Type IIA
orbifold, and odd in the case of Type IIB \cite{BG1}.

In the twisted \NSNS sector there exist fermionic zero modes along the
$n$ directions that are inverted by $\I_n$, and therefore the
condition only affects $s$. Let us introduce 
\be
\psi^\mu_{\pm} = {1 \over \sqrt{2}} 
\left( \psi^\mu_0 \pm i \tilde{\psi}^\mu_0 \right) \,,
\ee
where $\mu$ takes the $n$ values $\mu=9-n,\ldots,8$ and 
$\{\psi^\mu_0,\psi^\nu_0\}=\{\tilde\psi^\mu_0,\tilde\psi^\nu_0\}
=\delta^{\mu\nu}$; these modes satisfy then the Clifford algebra
\be
\{ \psi^\mu_{\pm} , \psi^\nu_{\pm} \} = 0\,, \qquad
\{ \psi^\mu_{+} , \psi^\nu_{-} \} = \delta^{\mu,\nu}\,.
\ee
The GSO-invariant boundary state is a linear combination of the two
states $\ket{B(r,s),\eta}_{\mbox{\scriptsize\NSNS,T}}$ with 
$\eta=\pm$. On the ground states $\ket{B(r,s),\eta}^0$, the fermionic
zero modes satisfy  
\be
\label{bound}
\begin{array}{ll}
\psi^\nu_\eta \ket{B(r,s),\eta}^0_{\mbox{\scriptsize\NSNS,T}} = 0 &
\hbox{for the Neumann directions $\nu=9-n,\ldots,8-n+s$,} \\
\psi^\nu_{-\eta} \ket{B(r,s),\eta}^0_{\mbox{\scriptsize\NSNS,T}} = 0 & 
\hbox{for the Dirichlet directions $\nu=9-n+s,\ldots,8$,}
\end{array}
\ee
where we have suppressed the dependence of the ground state on $k$
that is irrelevant for the present discussion. We may choose the
relative normalisation between the ground states corresponding to
$\eta=\pm$ to be defined by 
\be
\ket{B(r,s),+}^0_{\mbox{\scriptsize\NSNS,T}} 
= a \prod_{\nu=9-n+s}^{8} \psi^\nu_{-}
\prod_{\nu=9-n}^{8-n+s} \psi^\nu_{+} 
\ket{B(r,s),-}^0_{\mbox{\scriptsize\NSNS,T}} \,.
\ee
It then follows that 
\be
\ket{B(r,s),-}^0_{\mbox{\scriptsize\NSNS,T}} = 
b \prod_{\nu=9-n+s}^{8} \psi^\nu_{+}
\prod_{\nu=9-n}^{8-n+s} \psi^\nu_{-} 
\ket{B(r,s),+}^0_{\mbox{\scriptsize\NSNS,T}} \,,
\ee
where
\be\label{a1}
b = {(-1)^{n(n-1)/2} \over a} \,.
\ee
The expression $(-1)^{n(n-1)/2}$ is $+1$ for $n=0$ mod(4), and equals
$-1$ for $n=2$ mod(4). On the ground states the two GSO-projections
$(-1)^F$ and $(-1)^{\tilde{F}}$ take the form  
\be\label{GSOzero}
\begin{array}{rl}
(-1)^F & = c \prod_{\mu=9-n}^{8} (\sqrt{2} \psi^\mu_0) 
 = c \prod_{\mu=9-n}^{8} (\psi^\mu_+ + \psi^\mu_-) \,, \\
(-1)^{\tilde{F}} & = d \prod_{\mu=9-n}^{8} 
(\sqrt{2} \tilde{\psi}^\mu_0) 
= d i^n \prod_{\mu=9-n}^{8} (\psi^\mu_+ - \psi^\mu_-) \,.
\end{array}
\ee
Since both $(-1)^F$ and $(-1)^{\tilde{F}}$ have to be of order 
$2$, $c$ and $d$ satisfy
\be\label{a2}
c^2 = d^2 = (-1)^{n(n-1)/2} \,.
\ee
Applying (\ref{GSOzero}) to the boundary states we then find
\be
\begin{array}{rcl}
(-1)^F \ket{B(r,s),-}_{\mbox{\scriptsize\NSNS,T}} & = & 
{c\over a} \ket{B(r,s),+}_{\mbox{\scriptsize\NSNS,T}}\,, \\
(-1)^F \ket{B(r,s),+}_{\mbox{\scriptsize\NSNS,T}} & = & 
{c\over b} \ket{B(r,s),-}_{\mbox{\scriptsize\NSNS,T}}\,,  \\
(-1)^{\tilde{F}} \ket{B(r,s),-}_{\mbox{\scriptsize\NSNS,T}} & = & 
i^n {d\over a} (-1)^{n-s} 
\ket{B(r,s),+}_{\mbox{\scriptsize\NSNS,T}}\,, \\
(-1)^{\tilde{F}} \ket{B(r,s),+}_{\mbox{\scriptsize\NSNS,T}} & = & 
i^n {d\over b} (-1)^{s} 
\ket{B(r,s),-}_{\mbox{\scriptsize\NSNS,T}} \,.
\end{array}
\ee
We may choose for convenience 
$$ 
a = b = c = (-1)^{n(n-1)/4}\,, \qquad
i^n d = \kappa c \,,
$$ 
where $\kappa=\pm 1$; this is then consistent with (\ref{a1}) and 
(\ref{a2}). 

If the orbifold does not involve $(-1)^{F_L}$, \ie\ if it is a $g_1$
orbifold, then the left- and right GSO-projections have to be the same,
and thus a GSO-invariant combination only exists if 
$\kappa (-1)^s=+1$. In the case of a $g_2$ orbifold the situation
is precisely opposite, \ie\ a GSO-invariant boundary state exists
provided that $\kappa (-1)^s=-1$. In either case, the GSO-invariant
boundary state is then, up to normalisation,  given by
\be
\ket{B(r,s)}_{\mbox{\scriptsize\NSNS,T}}=
\ket{B(r,s),+}_{\mbox{\scriptsize\NSNS,T}}+
\ket{B(r,s),-}_{\mbox{\scriptsize\NSNS,T}}\,.
\ee

The analysis in the twisted \RR sector is very similar; in this case 
the fermionic zero modes occur for $\mu=1,\ldots,8-n$, and the
condition therefore only involves $r$. We can similarly introduce
modes $\psi^\mu_{\pm}$ that satisfy a Clifford algebra, and the 
boundary states $\ket{B(r,s),\eta}_{\mbox{\scriptsize\RR,T}}$ then
satisfy 
\be
\label{bound1}
\begin{array}{ll}
\psi^\nu_\eta \ket{B(r,s),\eta}^0_{\mbox{\scriptsize\RR,T}} = 0 & 
\hbox{for the Neumann directions $\nu=1,\ldots,r+1$,} \\
\psi^\nu_{-\eta} \ket{B(r,s),\eta}^0_{\mbox{\scriptsize\RR,T}} = 0 & 
\hbox{for the Dirichlet directions $\nu=r+2,\ldots,8-n$,}
\end{array}
\ee
where the suffix $0$ again denotes the ground state. We can
choose the relative normalisation of the ground states as
\be
\ket{B(r,s),+}^0_{\mbox{\scriptsize\RR,T}} = \hat{a}
\prod_{\nu=1}^{r+1} \psi^\nu_{-}\prod_{\nu=r+2}^{8-n} \psi^\nu_{+} 
\ket{B(r,s),-}^0_{\mbox{\scriptsize\RR,T}} \,,
\ee
and 
\be
\ket{B(r,s),-}^0_{\mbox{\scriptsize\RR,T}} = \hat{b}
\prod_{\nu=1}^{r+1} \psi^\nu_{+}
\prod_{\nu=r+2}^{8-n} \psi^\nu_{-} 
\ket{B(r,s),+}^0_{\mbox{\scriptsize\RR,T}} \,,
\ee
where
\be
\hat{a} \hat{b} = (-1)^{(8-n)(7-n)/2}\,.
\ee
On the ground states the two GSO-projections $(-1)^F$ and
$(-1)^{\tilde{F}}$ take the form  
\be\label{GSOzeroRR}
\begin{array}{rl}
(-1)^F & = \hat{c} \prod_{\mu=1}^{8-n} (\sqrt{2} \psi^\mu_0) 
 = \hat{c} \prod_{\mu=1}^{8-n} (\psi^\mu_+ + \psi^\mu_-) \,, \\
(-1)^{\tilde{F}} & = \hat{d} \prod_{\mu=1}^{8-n} 
(\sqrt{2} \tilde{\psi}^\mu_0) 
= \hat{d} i^n \prod_{\mu=1}^{8-n} (\psi^\mu_+ - \psi^\mu_-) \,,
\end{array}
\ee
and since $(-1)^F$ and $(-1)^{\tilde{F}}$ have to be of order $2$,
\be
\hat{c}^2=\hat{d}^2=(-1)^{(8-n)(7-n)/2}\,.
\ee
We thus find
\be
\begin{array}{rcl}
(-1)^F \ket{B(r,s),-}_{\mbox{\scriptsize\RR,T}} & = & 
{\hat{c}\over \hat{a}} \ket{B(r,s),+}_{\mbox{\scriptsize\RR,T}}\,, \\
(-1)^F \ket{B(r,s),+}_{\mbox{\scriptsize\RR,T}} & = & 
{\hat{c}\over \hat{b}} \ket{B(r,s),-}_{\mbox{\scriptsize\RR,T}}\,, \\
(-1)^{\tilde{F}} \ket{B(r,s),-}_{\mbox{\scriptsize\RR,T}} & = & 
i^n {\hat{d}\over \hat{a}} (-1)^{7-n-r} 
\ket{B(r,s),+}_{\mbox{\scriptsize\RR,T}}\,, \\
(-1)^{\tilde{F}} \ket{B(r,s),+}_{\mbox{\scriptsize\RR,T}} & = & 
i^n {\hat{d}\over \hat{b}} (-1)^{r+1} 
\ket{B(r,s),-}_{\mbox{\scriptsize\RR,T}} \,.
\end{array}
\ee
As before we may choose for convenience 
$$
\hat{a} = \hat{b} = \hat{c} = (-1)^{(8-n)(7-n)/4}\,,
\qquad
i^n \hat{d} = \hat\kappa \hat{c} \,,
$$
where $\hat\kappa$ is again $\pm 1$. In the case of a $g_1$ orbifold
of Type IIB or a $g_2$ orbifold of Type IIA, the left- and
right-GSO-projections are the same, and therefore a GSO-invariant
boundary state exists provided that $\hat\kappa (-1)^{r+1}=+1$;
similarly a GSO-invariant boundary state exists in the other two cases 
provided that $\hat\kappa (-1)^{r+1}=-1$; in either case, the boundary
state for the twisted \RR sector is then, up to normalisation, given
by 
\be
\ket{B(r,s)}_{\mbox{\scriptsize\RR,T}}=
\ket{B(r,s),+}_{\mbox{\scriptsize\RR,T}}+
\ket{B(r,s),-}_{\mbox{\scriptsize\RR,T}}.
\ee
In summary the following boundary states are thus GSO-invariant  
\be
\begin{array}{ll}
\ket{B(r,s)}_{\mbox{\scriptsize\NSNS}} & \hbox{for all $(r,s)$,} \\
\ket{B(r,s)}_{\mbox{\scriptsize\RR}} & \hbox{if $r+s$ is} \quad \left\{
\begin{array}{ll}
\hbox{even:} & \hbox{IIA-orbifold,} \\
\hbox{odd:} & \hbox{IIB-orbifold,}
\end{array} \right. \\
\ket{B(r,s)}_{\mbox{\scriptsize\NSNS,T}} & 
\hbox{if $\kappa (-1)^s$ is} 
\quad \left\{
\begin{array}{ll}
\hbox{$+1$:} & \hbox{$g_1$-orbifold,} \\
\hbox{$-1$:} & \hbox{$g_2$-orbifold,}
\end{array}\right. \\[2pt]
\ket{B(r,s)}_{\mbox{\scriptsize\RR,T}} & 
\hbox{if $\hat\kappa (-1)^r$ is} \quad \left\{
\begin{array}{ll}
\hbox{$+1$:} & \hbox{$g_1$-orbifold of IIA or $g_2$-orbifold of IIB,} \\
\hbox{$-1$:} & \hbox{$g_2$-orbifold of IIA or $g_1$-orbifold of IIB.} \\
\end{array} \right.
\end{array}
\ee
It is reasonable to assume that every theory possesses at least one
fractional brane; if we make this assumption, it follows that $\kappa$
and $\hat\kappa$ must be the same. For example, for the $g_1$
orbifold of IIA, a GSO-invariant boundary state exists in all sectors
provided that $(-1)^{r+s}$, $\kappa (-1)^s$ and $\hat\kappa (-1)^r$
are all $+1$; this requires that $\kappa\hat\kappa=+1$. The other
cases are similar. Actually, the value of $\kappa$ is completely
determined by this assumption once we consider the conditions that
come from the requirement that the boundary states must also be
invariant under the orbifold projection. Again, the boundary state in 
the untwisted \NSNS sector is invariant under both $g_1$ or $g_2$, but a 
non-trivial condition arises in the untwisted \RR sector. Indeed,
since there are fermionic zero modes, $\I_n$ acts on the corresponding
ground states (that are analogously defined to (\ref{bound}) and
(\ref{bound1})) as 
\begin{eqnarray}
\I_n \ket{B(r,s),+}^0_{\mbox{\scriptsize\RR}} & = & 
\prod_{\mu=9-n}^{8} (\sqrt{2}\psi^\mu_0) \prod_{\mu=9-n}^{8} 
(\sqrt{2} \tilde\psi^\mu_0)
\ket{B(r,s),+}^0_{\mbox{\scriptsize\RR}} \\
& = &  i^n (-1)^{n(n-1)/2} (-1)^s 
\ket{B(r,s),+}^0_{\mbox{\scriptsize\RR}} \,,
\end{eqnarray}
and similarly 
\be
\I_n \ket{B(r,s),-}^0_{\mbox{\scriptsize\RR}} = 
i^n (-1)^{n(n-1)/2} (-1)^{n-s} 
\ket{B(r,s),-}^0_{\mbox{\scriptsize\RR}}\,.
\ee
In the case of the $g_1$ orbifold, the boundary state is invariant
under the orbifold projection provided that $s$ is even; in the
case of a $g_2$ orbifold the condition is that $s$ is odd. If this is
to be consistent with what we found above, we have to choose
$\kappa=\hat\kappa=+1$. Incidentally, this is the convention for which
\be
(-1)^F (-1)^{\tilde{F}} = \I_n\,,
\ee
on the ground states of the twisted \NSNS sector.  

The condition in the twisted sectors is more difficult to analyse
since the definition of $\I_n$ in the twisted sector is {\it a priori}
ambiguous. The correct prescription seems to be that the physical
states in both twisted sectors have to have eigenvalue $+1$ ($-1$) 
with respect to the standard orbifold projection if the orbifold is of
type $g=g_1$ ($g=g_2$).\footnote{This is well known in the case of the
orbifold of Type IIB by $\I_4 (-1)^{F_L}$ \cite{BG2} where it follows
from considerations of supersymmetry.} This does not give any further 
restrictions for the states in the twisted \RR sector (since the
action of $\I_n$ on the ground states does not involve any fermionic
zero modes), and in the twisted \NSNS sector, we find
\begin{eqnarray}
\I_n \ket{B(r,s),+}_{\mbox{\scriptsize\NSNS,T}} & = & \prod_{\mu=9-n}^{8}
(\sqrt{2}\psi^\mu_0) \prod_{\mu=9-n}^{8} (\sqrt{2} \tilde\psi^\mu_0)
\ket{B(r,s),+}_{\mbox{\scriptsize\NSNS,T}} \\
& = & i^n (-1)^{n(n-1)/2} (-1)^{n-s} 
\ket{B(r,s),+}_{\mbox{\scriptsize\NSNS,T}} \,,
\end{eqnarray}
and similarly 
\be
\I_n\ket{B(r,s),-}_{\mbox{\scriptsize\NSNS,T}} = 
i^n (-1)^{n(n-1)/2} (-1)^{s} \ket{B(r,s),-}_{\mbox{\scriptsize\NSNS,T}}\,.
\ee
Thus $s$ has to be even if $g=g_1$, and odd if $g=g_2$; this
reproduces precisely the above constraints. Taking everything
together, the allowed boundary states are then given as in the main
part of the paper.

\section{K-theory versus Cohomology}
\setcounter{equation}{0}

As was pointed out in Section~\ref{sect4}, the lattice of D-brane
charges that is described by K-theory is a sublattice of maximal rank
of the (suitably normalised) cohomology lattice. This is something one
may expect on general grounds: K-theory and cohomology are equivalent
over the rational numbers (see for example \cite{mm}), and over the
integers, they can therefore only differ by a finite torsion group. In
this appendix we shall illustrate this difference between K-theory and
cohomology by working out the simplest case, the $\I_2(-1)^{F_L}$
orbifold, in detail. We shall also give the corresponding results for
$\I_2$ and for the two theories with $n=4$.

For $n=2$ mod $4$, the orbifold in question is a $\Zop_2$ orbifold of
Type 0A/0B, and therefore the the D-brane spectrum (\ie\ the K-theory
group), as well as the R-R spectrum (\ie\ the relevant cohomology
group) is doubled; for simplicity we shall only consider one copy,
\ie\ we shall pretend that the theory is really a $\Zop_2$ orbifold of
Type IIA/IIB. The relevant K-theory group for the case of the
$\I_2(-1)^{F_L}$ orbifold is then $\Zop^{\oplus 6}$ if $r$ is even
(odd) in Type B (A), and it is trivial otherwise. The non-trivial
elements arise from fractional branes that exist for $s=1$, and
truncated branes that exist for $s=0$ and $s=2$. For fixed $r$, the
cohomology charges can be described by a six-component vector, whose
first two entries correspond to the charge with respect to the
untwisted \RR form $C^{(r+2)}_{\mu_1\ldots \mu_{r+1}7}$ and
$C^{(r+2)}_{\mu_1\ldots \mu_{r+1}8}$. The remaining four entries
describe the charge with respect to the four twisted R-R forms. We
choose the normalisation so that the minimal charges are all integer;
then a  \dh$(r,0)$-brane has charges 
\be
(0,0;\pm 2,0,0,0)\,,
\ee
and the fractional D$(r,1)$-branes are described by
\be
(\pm 1,0;\pm 1,\pm 1,0,0),\spc (\pm 1,0;0,0,\pm 1,\pm 1)\,,\spc
(0,\pm 1;\pm 1,0,\pm 1,0),\spc (0,\pm 1;0,\pm 1,0,\pm 1)\,,
\ee
where all possible sign combinations are allowed. Finally the
\dh$(r,2)$-branes have the charges
\be
(0,0;\pm 1,\pm 1,\pm 1,\pm 1)\,,
\ee
where the number of $+$-signs is even. Let us denote by $\Lambda_K$
the lattice that is generated by these D-branes, and by $\Lambda_H$
the cohomology lattice, {\it i.e.} the lattice generated by the basis
vectors 
\be
e_i=(0,\dots,0,1,0,\dots,0)\,, \qquad i=1,\dots,6\,,
\ee
where the $1$ is placed in the $i$-th position. Clearly $\Lambda_K$ is
a sublattice of $\Lambda_H$, and they differ by the finite (torsion)
group $\Lambda_H/\Lambda_K$. 

The group $\Lambda_H/\Lambda_K$ is generated by the elements $e_3,e_4$
and $e_5$ since
\ba
e_1&=&(1,0;-1,-1,0,0)+e_3+e_4 \,,\\
e_2&=&(0,1;-1,0,-1,0)+e_3+e_5 \,,\\
e_6&=&(0,0;1,1,1,1)-e_3-e_4-e_5\,.
\ea
Each of these three elements is of order two since $2e_i$ (for
$i=3,4,5$) corresponds to a truncated  \dh(r,0)-brane and is hence in
$\Lambda_K$. Finally, none of the combinations  
\be
e_3+e_4,\;\;e_3+e_5,\;\;e_4+e_5,\;\;e_3+e_4+e_5
\ee
are elements of $\Lambda_K$, and thus
\be
\Lambda_H/\Lambda_K=\Zop_2\times\Zop_2\times\Zop_2\,.
\ee

In the case of the $\I_2$ orbifold, the six relevant \RR charges are
the untwisted \RR form $C^{(r+1)}_{\mu_1\ldots\mu_{r+1}}$ and 
 $C^{(r+3)}_{\mu_1\ldots\mu_{r+1}78}$, as well as the four twisted
\RR forms. Choosing again the normalisation of the charges so that
every D-brane has integer components, the fractional D$(r,0)$-brane is
described by
\be
(\pm 1,0;\pm 2,0,0,0)\,,
\ee
where all sign choices are allowed, and the 2 can be placed in any of
the four last entries. The \dh$(r,1)$-branes have charges
\be
(0,0;\pm 2,\pm 2,0,0),\spc (0,0;0,0,\pm 2,\pm 2),\spc
(0,0;\pm 2,0,\pm 2,0),\spc (0,0;0,\pm 2,0,\pm 2)\,,
\ee
whilst the fractional D$(r,2)$-branes are described by
\be
(0,\pm 1;\pm 1,\pm 1,\pm 1,\pm 1)\,.
\ee
Here any combination of signs with an even number of plus signs in the
last four entries is allowed. In this case the group
$\Lambda_H/\Lambda_K$ is generated by $e_3,e_4-e_3,e_5-e_3,e_6-e_3$,
where $e_3$ is of order four while each of the other three elements is
of order two in $\Lambda_H/\Lambda_K$. There are no further relations, 
and the group is therefore 
\be
\Lambda_H/\Lambda_K=\Zop_4\times\Zop_2\times\Zop_2\times\Zop_2\,.
\ee

We have also determined the torsion groups for the two orbifolds with
$n=4$\footnote{We have used Mathematica to obtain the following
results, in particular the function {\sf LatticeReduce} and the
Elementary Decomposition package.}. For the $\I_4$ orbifold the
torsion group is  
\be
\Lambda_H/\Lambda_K=\Zop_8\times(\Zop_4)^5\times(\Zop_2)^{10}\,,
\ee
while in the case of $\I_4 (-1)^{F_L}$ the answer is
\be
\Lambda_H/\Lambda_K=(\Zop_4)^5\times(\Zop_2)^{10}\,.
\ee

\section*{Acknowledgements}

We are grateful to O. Bergman, R. Blumenhagen, R. Dijkgraaf,
M.B. Green, F. Quevedo, A. Sen, C.T. Snydal, C.B. Thomas and in
particular to G.B. Segal for useful discussions. \\
M.R.G. is supported by a College Lectureship of Fitzwilliam College,
Cambridge.

\end{document}